\documentclass[11pt,a4paper]{article}
\usepackage{jcappub}
\usepackage{amsmath}
\usepackage{amssymb}
\usepackage{mathtools}

\newcommand{\e}{{\mathrm{e}}}

\renewcommand{\d}{\partial}
\renewcommand{\l}{\left(}
\renewcommand{\r}{\right)}
\newcommand{\la}{\langle }
\newcommand{\ra}{\rangle }
\newcommand{\be}{\begin{equation}}
\newcommand{\ee}{\end{equation}}
\newcommand{\ba}{\begin{align}}
\newcommand{\ea}{\end{align}}
\newcommand{\bg}{\begin{gather}}
\newcommand{\eg}{\end{gather}}
\newcommand{\bseq}{\begin{subequations}}
	\newcommand{\eseq}{\end{subequations}}

\newcommand{\mbf}{ \mathbf}

\newcommand{\erf}{ \e^{\sqrt\frac{2}{3} \frac{\phi}{M_P}} }

\begin{document}
	\title{A heatwave affair: mixed Higgs-\boldmath$R^2$ preheating on the lattice}

    \author{Fedor Bezrukov}
    \author{Chris Shepherd}
    
    \affiliation{School of Physics and Astronomy, The University of Manchester  \\ Manchester M13 9PL, United Kingdom}
    
    \emailAdd{Fedor.Bezrukov@manchester.ac.uk}
    \emailAdd{christopher.shepherd-3@postgrad.manchester.ac.uk}
        
\abstract{
	We use lattice methods to perform the first nonlinear study of preheating in $R^2$-healed Higgs inflation for ``$R^2$-like'' parameters $1.1\times 10^9$ and $1.8\times 10^9$ where the curvature-squared coupling $\beta$ and nonminimal coupling $\xi$ of the Higgs field contribute similarly to the CMB scalar perturbations. Preheating occurs first through tachyonic production of Higgs bosons, and later scattering off the homogeneous inflaton field. We generalise our results to ``Higgs-like'' parameters with smaller $\beta$, where observables saturate the bound of instantaneous preheating. All predictions for the spectral index and tensor-to-scalar ratio lie within the $1\sigma$ region of measurements by the Planck satellite, but a future ground-based experiment optimised for 21 cm tomography may be able to discriminate the mixed Higgs-curvature inflation from the pure Higgs and $R^2$ theories. 
}
\maketitle
\flushbottom
\section{Introduction}
\label{sec:intro}
Nonminimal Higgs inflation occupies a special position among the range of viable inflationary theories. In addition to solving the horizon and flatness problems and predicting a primordial scalar tilt and tensor-to-scalar ratio within experimental bounds, it does so without introducing any new degrees of freedom between the electroweak and Planck scales \cite{Bezrukov_2008}. Inflation is driven by the standard model (SM) Higgs, which has been discovered and studied extensively at the LHC \cite{Chatrchyan_2012}.

However, a new scale appears in the model due to the nonminimal Higgs-curvature coupling, which is a dimension 5 operator \cite{Burgess:2009ea,Barbon:2009ya,Burgess:2010zq,Hertzberg:2010dc}. It can be argued that this scale does not influence inflationary dynamics, and is virtually undetectable in present day experiments \cite{BezrukovSibiryakovMagnin}, as far as the scale of tree level unitarity violation depends on the homogeneous background fields. However, during preheating particles are produced with momenta exceeding this scale \cite{Ema_Higgs}, meaning that the reheating dynamics are essentially strongly coupled. This makes simple perturbative or semiclassical methods applied to reheating in the pure Higgs inflation model impossible to control \cite{Bezrukov_2009,GarciaBellidoRubio,PhysRevD.99.083519}.

The viable way to approach this problem is to study instead a UV complete model that would correspond to Higgs inflation both at low energies and during inflation.  The perturbative UV completions form the simple class of such models that can be fully analysed within standard QFT methods, which achieve the UV completion by addition of a new degree of freedom with mass below the problematic scale \cite{Giudice:2010ka,Ema_2017,Gorbunov_2019}. Here we focus on the model of \cite{Ema_2017}, where this degree of freedom corresponds to the scalaron particle appearing if $R^2$ is present in the Jordan frame action.
The model is parameterised by the coefficient $\beta$ in front of the $R^2$ term. For smaller $\beta$ the scalaron degree of freedom provides UV completion of the model, without altering the inflationary dynamics significantly (``Higgs-like'' parameters), while for larger $\beta$ the inflation becomes similar to the original Starobinsky inflationary model \cite{Starobinsky:1979ty} (``$R^2$-like'').

The inflationary dynamics of this theory have been studied extensively \cite{Ema_2017,gundhi2018scalaronhiggs}, but the details of its reheating are yet to be robustly determined. \cite{He_2019} found that the peak-like feature in the mass of the Nambu-Goldstone modes of the Higgs field is not sufficiently pronounced to preheat the Universe after one scalaron oscillation. \cite{Bezrukov:2019ylq} recently found that for special values of parameters, inhomogeneous modes were amplified at an exponential rate due to a tachyonic instability in the Higgs mass. It was tentatively proposed that the preheating in this case is cosmologically instantaneous, and that this result may be generalised to the full theory parameter space. However, these conjectures were made using a linearised treatment that did not account for backreaction, and could therefore not be taken verbatim. 

In this paper we use semiclassical methods to study the self-consistent dynamics of preheating numerically, for the $R^2$-like parameters  $1.1\times 10^9 \lesssim\beta \lesssim 1.8\times10^9$ in the scalar and Goldstone sectors.  This range captures the typical ``mixed'' preheating dynamics that apply right down to the lower limit for $\beta$, and is close to the pure $R^2$ limit. At the same time it is sufficiently far from the strongly coupled HI limit to allow for a successful semiclassical treatment. We find that the mixed preheating can be split into two stages: an initial stage of tachyonic production, followed by a longer stage of scattering off the scalaron condensate along with freely-turbulent cascade of spectra towards the UV. The tachyonic stage occurs regardless of whether dynamics are critical at the first scalaron zero crossing and lasts 1--2 e-foldings. The rescattering stage becomes insufficient to fully destroy the homogeneous scalaron for parameters closer to $R^2$-like limit, and the remaining scalaron condensate must decay perturbatively. For our parameters, we find that the pivot scale exits the horizon  between $57.8$ and $58.4$ e-foldings before the end of inflation, for the upper and lower limits of $\beta$ in our range respectively, allowing one to predict the scalar spectral index $n_s$ and the tensor-to-scalar ratio $r$. The value of $N_e$ for our lower limit of $\beta$ nearly saturates the bound of instantaneous preheating, and for Higgs-like parameters $\beta<10^9$ we expect preheating to be even faster --- with little effect on observables.

This paper is structured as follows. In section \ref{sec:background_dynamics} we review the homogeneous field dynamics and in section \ref{sec:inhomogeneous_dynamics} we outline analytically the approximate dynamics of inhomogeneous fluctuations that preheat the Universe. In section \ref{sec:method} we summarise the numerical method of our simulations and specify the connection between our results and cosmological observables. Our results are summarised in section \ref{sec:results}, and their implications and validity are discussed in section \ref{sec:discussion}.

\section{Background field dynamics}
\label{sec:background_dynamics}
The action in the scalar sector for $R^2$-modified nonminimal Higgs inflation in the Jordan frame is \cite{Ema_2017}
\be
\label{action-1}
S_J=\int d^4x \sqrt{-g} \left[- \l\frac{M_P^2}{2}+\xi H^\dagger H \r R+\frac{\beta}{4}
R^2+ g^{\mu\nu}\d_{\mu}H^\dagger\d_\nu H-\lambda \l H^\dagger H \r^2\right],
\ee
where $g_{\mu \nu}$ is the metric in FRW spacetime with scale factor $a(t)$, $\l 1, -a(t), -a(t), -a(t) \r$, and $g$ is its determinant. $R$ is the Ricci scalar and $H$ is the standard model Higgs doublet. $M_P$ is the reduced Planck mass, related to the Planck mass $M_{Pl}$ via $M_P \equiv M_{Pl}/(8\pi)$. The present-day Higgs vacuum expectation value $\approx 246$ GeV is far below inflationary and preheating scales, and we omit it throughout.

For the remainder of our discussion, we transform into the Einstein frame with the conformal transformation $g_{\mu \nu} \rightarrow \tilde{g}_{\mu \nu}\equiv \e^{\sqrt{\frac{2}{3}}\frac{\phi}{M_P}} g_{\mu \nu}$ (see \cite{Ema_2017} for details). Dropping tildes for convenience and referring to $|h|\equiv \sqrt{ 2 H^\dagger H}$ as the radial Higgs degree of freedom, one obtains the action
\be \label{scalarAction}
\begin{split}
	S=&\int d^4 x \sqrt{-g}\, \Biggl[ -\frac{M_P^2}{2}R +
	\frac{1}{2}\e^{-\sqrt{\frac{2}{3}}\frac{\phi}{M_P}}
	g^{\mu\nu}\d_{\mu} |h| \d^\mu |h| + g^{\mu\nu}\d_{\mu}\phi\d_\nu \phi\\
	 &	-\frac{1}{4}\e^{-2 \sqrt{\frac{2}{3}}\frac{\phi}{M_P}}\l\lambda
	|h|^4 +\frac{M_P^4}{\beta}\l \e^{\sqrt{\frac{2}{3}}\frac{\phi}{M_P}}-1-\xi \frac{|h|^2}{M_P^2}\r^2\r\Biggr].
\end{split}
\ee
The curvature-squared term provides an additional degree of freedom, which manifests itself in the Einstein frame as the ``scalaron'' field $\phi$. It was observed in \cite{Gorbunov_2019} that the theory is perturbative if
\be \label{conditon:perturbativity}
	\beta \gtrsim \frac{\xi^2}{4\pi}.
\ee
We parametrise the Higgs and Goldstone degrees of freedom using four real fields,
\begin{equation}
h = \begin{pmatrix} h_1 \\ h_2 \\ h_3 \\ h_4 \end{pmatrix},
\end{equation}
where $|h| = \sqrt{h_1^2 + h_2^2 + h_3^2 + h_4^2}$. This parametrisation has a global rotational symmetry in field space.

For positive scalaron values, the Higgs field has the family of potential minima,
\begin{equation}
|h|^2 = \frac{\xi M_P^2}{\xi^2 + \lambda \beta} \l \erf -1 \r, 
\end{equation}
and the potential along this special direction is
\be
	V_{\mathrm{inf}} (\phi) = \frac{\lambda M_P ^4}{4 \l \xi^2 + \lambda \beta \r } \l 1 - e^{- \sqrt\frac{2}{3} \frac{\phi}{M_P}} \r ^2,
\ee
which is precisely the potential of nonminimal Higgs inflation with the substitution $\xi^2 \rightarrow \l \xi^2+ \lambda \beta\r$. It is therefore straightforward to generalise the dynamics of pure Higgs inflation to the mixed case. Inflation proceeds with initially super-Planckian homogeneous $\phi$, and isocurvature modes are suppressed by expansion \cite{Ema_2017,gundhi2018scalaronhiggs}. We denote the quantum expectation value for fluctuations of $\phi$ and $h$, on top of the uniform inflationary background, as the homogeneous fields $\left<\phi\right>\equiv\phi_{(0)}$ and $\left<|h|\right>\equiv h_{(0)}$ respectively. During inflation, the inflaton field is sufficiently homogeneous that one may regard $\phi_{(0)}$ and $h_{(0)}$ as classical backgrounds and quantise small spatially-dependent fluctuations on top. As fluctuations are enhanced to large occupation values during preheating, a semiclassical treatment becomes appropriate \cite{classical_decay}. The normalisation of the scalar CMB power spectrum \cite{Akrami:2018odb} gives the constraint
\be \label{specNorm}
	\beta+\frac{\xi^2}{\lambda}\approx 2\times 10^{9}.  
\ee
The upper limit suggested in \cite{Gorbunov_2019},
\be\label{conditon:CMB}
	\beta \lesssim \frac{\xi^2}{\lambda},
	\ee 
corresponds to the Higgs nonminimal coupling giving the leading contribution to the CMB normalisation. We therefore classify parameters that satisfy \ref{conditon:CMB} as ``Higgs-like'' and those with larger $\beta$ that do not as ``$R^2$-like'' (c.f.\ Figure \ref{fig:parameter_space}).

Inflation ends when the slow roll parameter,
\be
	\eta = \left| \frac{\ddot{\phi}_{(0)}}{3 \mathcal{H} \dot{\phi}} \right|,
\ee
where a dot denotes a derivative with respect to physical time and $\mathcal{H}$ is the Hubble rate, equals one and the scalaron begins to oscillate. After this momement, $\phi_{(0)} < 0.1 M_P$ and $h_{(0)} \sim \l\frac{\xi}{\xi^2 + \lambda \beta} \, \phi_{(0)}\, M_P\r^{1/2}$, so for our analytic discussion we may simplify the scalar potential:
\be
\begin{split} \label{simplePotential}
	V(\phi,|h|) & =\frac{1}{4}\l \lambda +\frac{\xi^2}{\beta}\r  |h|^4
	+ \frac{M_P^2}{6\beta}\phi^2 - \frac{\xi M_P}{\sqrt{6}\beta}\phi |h|^2\, \\
	&+ \frac{7}{108\beta}\phi^4 -
	\frac{M_P}{3\sqrt{6}\beta}\phi^3  + \frac{\xi}{6\beta}\phi^2 |h|^2+ \ldots
\end{split}
\ee
Consider the second line: the first two terms are Planck-suppressed compared to the second term of the first line, so can be neglected for analytic estimates of preheating when comparing the potential for homogeneous fields. The last term of the second line is also Planck-suppressed compared to the last term on the first line. Furthermore, the smallness of $\phi_{(0)} / M_P$ ensures that the Hubble parameter is much smaller than the scalaron mass during all of preheating. One may therefore ignore expansion when discussing a small number of scalaron oscillations.

The homogeneous background evolution in the potential \eqref{simplePotential} was analysed in detail in \cite{Bezrukov:2019ylq}.  In this section we outline the modifications to this dynamics arising from the backreaction from the created particles. The full treatment can be obtained by using semiclassical approximation of \cite{PhysRevD.70.043538} which is valid when large numbers of particles are created.  Here we will estimate qualitatively the effects of the large number of generated Higgs perturbations on the particle production.  The main interest for us is the evolution of the inhomogeneous modes of the Higgs field, which gives the main contribution to the reheating \cite{Bezrukov:2019ylq}.

In the presence of the excitations of the Higgs field, the evolution of the Higgs mode with conformal momentum $\mbf{k}$ is
\begin{equation}
    \ddot{h}_{i,\mbf{k}}+3\mathcal{H}\dot{h}_{i,\mbf{k}} + \l \frac{k^2}{a^2}
    +m_{h,\mathrm{eff}}^2 \r h^i_\mbf{k}+\dots
    = 0,
\end{equation}
where $i$ can range from $1$ to $4$, $k\equiv |\mbf{k}|$, dots correspond to the scattering terms, and the effective mass is
\begin{equation} \label{mhDef}
	m_{h,\mathrm{eff}}^2 = -\sqrt\frac{2}{3} \frac{\xi}{\beta} M_P\, \phi_{(0)}
	+ \l \lambda + \frac{\xi^2}{\beta} \r 
	  \l 3|h_{(0)}|^2 + \la (h_i-\left<h_{i}\right>)^2 \ra \r.
\end{equation}
This mass becomes tachyonic for positive $\phi_{(0)}$ if $h_{(0)}$ is sufficiently small (see Figure \ref{trajectoryPlots}).  However, the created excitations of the Higgs field give positive contribution to $m_{h,\mathrm{eff}}$.
The last term gives the inhomogeneous contribution and is unimportant at early times, but dominates at late times, blocking the tachyonic production. Expression \eqref{mhDef} neglects the contribution to $	m_{h,\mathrm{eff}}$ from scalaron fluctuations, sourced by the final term of \eqref{simplePotential}, as far as this contribution is subleading to the final term at all times. Therefore, both the homogeneous and inhomogeneous terms in the last line of \eqref{simplePotential} are subleading to the second line, and we neglect the former throughout the rest of our analytic discussion.

During the intervals $\phi_{(0)}<0$, all terms in the first line of \eqref{simplePotential} are positive and the scalaron oscillates with frequency 
\be \label{mxDef}
	 {m_\phi^{(-)}}^2 = \frac{M_P^2}{6 \beta}.
\ee
Meanwhile the radial Higgs is oscillating about zero and the first term of \eqref{mhDef} is initially dominant. The $SU(2)$ symmetry of the theory is manifest during these intervals. While $\phi_{(0)}>0$, the scalaron oscillates with frequency 
\be \label{msDef}
	 {m_\phi^{(+)}}^2 = \frac{\lambda M_P^2}{3\l\xi^2 + \lambda \beta\r}.
\ee 
The effective potential now has a local maximum in the Higgs directions at $\left<h\right> =0$, and the family of degenerate minima
\be
h_{\mathrm{min}}^2(\phi_{(0)}) \equiv \sqrt\frac{2}{3} \frac{\xi}{\xi^2 + \lambda \beta} M_P\, \phi_{(0)}.
\ee
During these intervals homogeneous the Higgs will oscillate about this minimum. When $|h|$ acquires a vacuum expectation value, the gauge symmetry of the theory is broken and the three directions orthogonal to $|h|$ become Goldstone bosons. At late times, the nature of the symmetry breaking becomes somewhat less clear, as the constant $m_\phi^{(\pm)}$ and the field-dependent $	m_{h,\mathrm{eff}}$ become closer in magnitude. The scalaron amplitude drops with expansion proportional to $a^{-3/2}$, so the condition for highly separated Higgs and scalaron mass scales is
\be	\label{earlyTime}
 \frac{a(t)}{a_e}\ll \l\frac{\sqrt{6}\xi\phi_e }{M_P}\r^{2/3},
\ee
where the $e$ subscript denotes the end of inflation, and the scalaron amplitude $\phi_e \sim 0.1 M_P$. This condition strongly holds for multiple e-foldings for all our parameters, and only fails to hold for larger $\beta$ as one takes the $\xi\rightarrow 0$ limit of pure $R^2$ preheating. Away from this ``deep $R^2$-like'' parameter range, the qualitative features of the mixed preheating discussed here can therefore be generalised to all parameters of the mixed theory. 

When the scalaron background first crosses from negative to zero, the homogeneous Higgs may end up in either of the two potential valleys $\pm h_{\mathrm{min}} (\phi_{(0)})$. These valleys are of course connected in the full field space of the complex Higgs doublet, but the homogeneous dynamics at this time only concern one Higgs direction.  The choice of valley depends on the phase of the homogeneous Higgs oscillations at the time of scalaron crossing, i.e.\ on the ratio $	m_{h,\mathrm{eff}} / {m_\phi^{(-)}}$, which is determined by the parameters $\beta$ and $\xi$ and the value of $\dot\phi_{(0)}$. For special values of this phase, the homogeneous Higgs may spend a significant fraction of a scalaron period at the local maximum $h_{(0)} =0$ before finally rolling down to the minimum of the effective potential.  This scenario triggers a tachyonic instability in the Higgs mass, which enhances fluctuations during the first stage of preheating.

\section{Inhomogeneous modes}
\label{sec:inhomogeneous_dynamics}
We now turn to the dynamics of fluctuations on top of the homogeneous fields, whose enhancement destroys the inflaton and reheats the Universe. The situation for the scalaron is somewhat trivial, as its effective potential does not receive a significant contribution from Hartree-type terms, and depends only on the parameters $\lambda$, $\xi$ and $\beta$. The homogeneous and inhomogeneous scalaron modes therefore oscillate with the masses $m_\phi^{(+)}$ or $m_\phi^{(-)}$, depending on the sign of $\phi_{(0)}$ (c.f.\ Figure \ref{meansPlots}). When the homogeneous Higgs oscillates around the minimum of its effective potential, both the homogeneous and inhomogeneous fluctuations are expanded around zero or $h_{\mathrm{min}}(\phi_{(0)})$, again depending on the sign of $\phi_{(0)}$. 

However, the homogeneous and inhomogeneous Higgs modes behave differently during critical dynamics, when expanding \eqref{mhDef} around zero for a positive scalaron gives a negative square mass for all four inhomogeneous Higgs directions. This tachyonic instability triggers exponential growth of inhomogeneous Higgs modes. Enhanced fluctuations contribute to the third term of \eqref{mhDef} and may cancel the tachyonic behaviour much earlier than the homogeneous field dynamics alone. Eventually, sufficiently large fluctuations spoil the tuning required for critical dynamics and prevent this situation from repeating at any future time. Furthermore, when \eqref{earlyTime} is no longer satisfied, the scales $	m_{h,\mathrm{eff}}$ and ${m_\phi^{(+)}} < m_{h,\mathrm{eff}}$ become comparable in magnitude, reducing the tachyonic mass and making particle production less dramatic. One therefore would like to know whether an initially noncritical ratio $m_{h,\mathrm{eff}}/{m_\phi^{(-)}}$ gives significant tachyonic production before this mechanism turns off.

Once tachyonic production terminates, further depletion of the scalaron is achieved by scattering of a Higgs off the homogeneous scalaron, which we describe as ``rescattering''. This processes is mediated by the third term of \eqref{simplePotential} with the exchange of a virtual Higgs, and its cross section varies with $\beta$ as $\l \xi/ \beta\r^4$. Physical occupation numbers drop with expansion proportional to $a^{-3}$, and if the interaction rate drops to below the Hubble rate, rescattering shuts off. Inhomogeneous Higgs modes also scatter off one another via the first term of \eqref{simplePotential}, causing the spectrum to move towards the UV \cite{PhysRevD.70.043538}.

When rescattering terminates, the homogeneous scalaron may not have been completely destroyed. In this case, it persists until it decays perturbatively through the third term of \eqref{simplePotential}. In this regime, the Higgs oscillates too slowly to settle into $h_{\mathrm{min}}(\phi_{(0)})$, so the scalaron has mass $m_{\phi}^{(-)}$ and one expands \eqref{mhDef} around $h_{(0)} = 0$. The decay width of the scalaron into two radial Higgs is
\be \label{GammaPhi}
	\Gamma_\phi 	\approx  \frac{1}{24 \pi {m_\phi^{(-)}}}\l   M_P \frac{\xi}{ \beta} \r ^2    \sqrt{1 - 2 \frac{m_{h,\mathrm{eff}}^2}{{m_\phi^{(-)}}^2}}.
\ee
and the $m_{h,\mathrm{eff}} \rightarrow 0$ limit of $\Gamma_\phi$ gives the decay width into each Goldstone direction. Decays into two radial Higgs are kinematically allowed once $2 m_{h,\mathrm{eff}} < {m_\phi^{(+)}}$.  Later than our simulation run-time, one finds this condition to be satisfied before the total decay width becomes larger than the Hubble parameter and the scalaron efficiently decays.

\section{Numerical method} 
\label{sec:method}
In order to study the self-consistent dynamics of the system, we solve its classical equations of motion exactly in a discretised spacetime of spatial volume $L^3$ with $N^3 = 64^3$ points. We use a modified version of GABE \cite{Child:2013ria,gabe_ref}, whose second-order Runge-Kutta integrator can cope with single time derivatives in the equations of motion. These are unavoidable due to the non-canonical Higgs kinetic term in \eqref{scalarAction}. 

We solve the equations of motion,
\begin{align}\label{eom1}
 \ddot{\phi} - \frac{\nabla^2}{a^2} \phi + 3\frac{\dot{a}}{a}\dot{\phi} +  \frac{\partial V}{\partial \phi} + \frac{1}{\sqrt{6} M_P}e^{- \sqrt\frac{2}{3} \frac{\phi}{M_P}} \sum_i \l\dot{h}_i^2 - \l a^{-1}\mbf{\nabla} h_i  \r^2 \r &= 0, \\ \label{eom2}
	\ddot{h}_i - \frac{\nabla^2}{a^2} h_i + 3\frac{\dot{a}}{a} \dot{h}_i - \sqrt\frac{2}{3} \frac{1}{M_P} \l \dot{\phi}\dot{h}_i - a^{-2}\mbf{\nabla} \phi \cdot \mbf{\nabla} h_i \r
		+ e^{\sqrt\frac{2}{3} \frac{\phi}{M_P}} \frac{\partial V}{\partial h_i}&=0, 
\end{align}
\be \label{friedmann}
	\l \frac{\dot{a}}{a}\r^2 \equiv \mathcal{H}^2 = \frac{1}{3M_P ^2}\left<\rho\right>,  
\ee
in physical time, where
\be
	\rho = \frac{1}{2} \left[ \dot{\phi}^2 + a^{-2}\left( \mbf{\nabla} \phi \right)^2 + e^{-\sqrt\frac{2}{3} \frac{\phi}{M_P}} \left( \dot{h}_i^2 + a^{-2} \left( \mbf{\nabla} h_i \right) ^2 \right) \right] +V,
\ee
and we use $\left<\ldots\right>$ to denote a grid average when discussing our simulation. This average coincides with the quantum expectation value in the semiclassical limit, which is realised almost immediately. Here, $V$ is the exact potential of \eqref{scalarAction} rescaled by $m_\phi^{(+)-4}$. A value of $0.01$ is used for $\lambda$ throughout. The homogeneous fields $\phi_{(0)}$ and $h_{(0)}$ are initialised to their values at the end of inflation, with the $h_1$ direction chosen for the homogeneous Higgs, and the initial scale factor is set to unity. We consider approximately canonically-normalised fields 
\be
	f_0 = \phi,\quad f_i = e^{- \frac{\phi}{\sqrt{6} M_P}} h_i
\ee
and initialise fluctuations in Fourier space, using the rotated basis $f_{a} \rightarrow \tilde{f}_a$, $a=(0,1,2,3,4)$, that diagonalises
\be \label{mijDef}
M_{ab} \equiv \left< \frac{\partial^2 V}{\partial f_a \partial f_b} \right>.
\ee
Suppressing an ``a'' suffix, a Fourier mode $\tilde{f}_\mbf{k}$ with conformal momentum $\mbf{k}$ is initialised if $k>2 \pi / L$, where the latter is the infrared cutoff of the box. We sample from a Gaussian random distribution,
\be\label{fInit}
	\left< |\tilde{f}_\mbf{k}|^2 \right> = \frac{1}{2 a^3 \omega_k} 
\ee 
where
 \be \label{omegaDef}
	\omega_k^2 = \frac{k^2}{a^2} +  \left<\frac{\partial^2 V}{\partial \tilde{f} ^2}\right>,
\ee 
and the scale factor rescales the field into its conformal analogue. Different modes are initialised with random oscillatory phases, and the positive and negative frequency time derivatives are respectively initialised as
\be\label{dfInit}
	\dot{f}_\mbf{k} = \pm i \omega_k f_{\mbf{k}},
\ee
where we've ignored irrelevant Hubble-scale terms. The adiabatic invariant number density for a conformal mode $\mbf{k}$ is
\begin{equation}\label{nkDef}
n_k = \frac{a^3}{2 \omega_k} \left( |\dot{\tilde{f}}_\mbf{k}|^2 + \omega_k^2 |\tilde{f}_\mbf{k}|^2\right)
\end{equation}
so these initial conditions correspond to an average $n_k$ of $1/2$. Fluctuations are then rotated and rescaled back into the basis $(\phi, h_i)$, and added to the homogeneous field values in position space. It was found in \cite{Bezrukov:2019ylq} that the rotation step significantly modifies the initial amplitude of Higgs and Goldstone fluctuations.

The parameters of the lattice are chosen to contain all relevant scales for particle production, from the tachyonic scale up to the highest momentum scales of particles created during free turbulence. These must fit between the lattice IR and UV scales of $2\pi / L$ and $2 N \pi / L$ respectively. An additional complication arises due to the significant contribution of vacuum fluctuations in the UV to the energy of the system. However, these modes are only occupied due to rescattering and turbulent flow to the UV, so their initial occupation is irrelevant to final results. We therefore use an initial UV cutoff $\Lambda_{\mathrm{init}}$ above which modes are initialised with zero occupation, which is chosen to fully capture the initial particle creation, used by \cite{Repond_2016} for example. We use a time step of $L/20$, which captures the fastest oscillations in the box with sufficient accuracy that further reductions do not modify spectra.
 
  For $\beta\lesssim 10^9$, our initialisation procedure still gives a semiclassical zero-point energy comparable with the total energy. This is due to to the large ratio of the Higgs zero-point energy to the homogeneous energy in the scalaron,
 \be \label{zero_point_energy}
 \l \frac{{m_\phi^{(+)}}^2}{2} \phi_{(0)} ^2 \r^{-1}\int_{k^2<\mu^2} \frac{d^3 k}{\l 2 \pi \r^3}
        \frac{\sqrt{k^2 + m_{h,\mathrm{eff}}^2 } }{2}\sim 0.01 \,\frac{\xi^2}{\beta} \l 1 + \frac{\xi^2}{\lambda \beta}\r^{-1},
 \ee
 where $\mu^2$ is the maximally tachyonic mass due to the first term on the right hand side of \eqref{mhDef} alone. Preheating for these parameters therefore proceeds without attaining the semiclassical limit, and will instead occur via quantum processes which are not captured by our approach. The study of preheating in this case requires the use of the 2PI formalism \cite{Lindner:2005kv}, which is extremely computationally demanding if expansion of the universe is taken into account, so we perform our semiclassical simulations in the $R^2$-like region of parameter space $\beta >10^9$ where such an approach is not required. We will see in section \ref{sec:discussion} that the the preheating for our parameters is extremely fast, and cosmologically near-instantaneous for the lower end of our range for $\beta$. We expect no significant change in observables for smaller $\beta$, and may therefore generalise our results to Higgs-like parameters.
 
Zero-point fluctuations may also alter whether or not dynamics are initially critical, due to the third term in \eqref{mhDef}. This alteration is not physical in origin, as it arises due to semiclassical effects with $n_k \sim 1$ and does not approximate quantum modifications to the effective potential. We parameterise this effect by comparing the initial inhomogeneous contribution to \eqref{mhDef} to the difference in $m_{h,\mathrm{eff}}$ for adjacent critical parameters, $\beta_n$ and $\beta_{n+1}$. For these parameters, the homogeneous Higgs makes $n$ and $(n+1)$ oscillations respectively while $\phi_{(0)} <0$. Their difference in $m_{h,\mathrm{eff}}$ is
 \be
    m_{h,\mathrm{eff}}|_{\beta_{n+1}} - m_{h,\mathrm{eff}}|_{\beta_n} = \l \frac{1}{n} -\frac{1}{n+1}\r {m_\phi^{(-)}}
 \ee
 and we find the ratio of the zero-point contribution to $m_{h,\mathrm{eff}}$ over this difference --- and the detuning of initially-critical homogeneous dynamics --- to be of the order $10^{-2}$ for our simulations.

During the simulation, the inhomogeneous contributions to the kinetic, gradient and potential energy --- $ K(\phi, \dot{\phi}, \dot{h}_i)$, $ G (\phi, \vec{\nabla} \phi, \vec{\nabla} h_i)$ and $\left<V (\phi,|h|)\right>$ of \eqref{scalarAction} respectively --- are determined from the grid averages $\left< \phi \right>$, $\left<h_i\right>$ and their derivatives as usual,
\begin{equation}\label{kinDef}
K_{\mathrm{in}} = \left< K (\phi,\dot{\phi}, h_i) \right> - K(\left<\phi\right>,\dot{\left<\phi\right>}, \left<h_i\right>)
\end{equation}
and likewise for $G$, $V$ and $\rho=K+G+V$.  Note that scalar Hartree effective potential can be approximately written as
\be \label{minShift}
V_H(\phi, |h|) \approx \frac{{m_\phi^{(+)}}^2}{2} \phi_{(0)}^2+ \frac{1}{4}\l \lambda + \frac{\xi^2}{\beta} \r \l |h_{(0)}|^2+ \int_\mbf{k} |h_{\mbf{k}}|^2  - \frac{\xi}{\xi^2 + \lambda \beta} \sqrt\frac{2}{3} \phi_{(0)} M_P \r ^2,
\ee
which is minimised for $\phi_{(0)} >0$ by setting the second term to zero. Calculating $V(\left<\phi\right>, \left<|h|\right>)$ ignores the inhomogeneous contribution to this term \eqref{minShift} and gives an apparently negative $V_{\mathrm{in}}$. This highlights the inconsistency in calculating the homogeneous potential in an arrangement whose dynamics are governed by $V_H$, and $V_{\mathrm{in}}$ should not be trusted in this case. One can however trust $K_{\mathrm{in}}$ and $G_{\mathrm{in}}=G$ at all times, and when the rotational symmetry about $|h|=0$ is restored $V_{\mathrm{in}}$ can also be trusted.

In order to calculate particle occupation spectra, fields are transformed from $(\phi, h_i) \rightarrow \tilde{f}_a$. For a given canonical field $\tilde{f}$, the conformal and adiabatic invariant number density is calculated using \eqref{nkDef}. The energy density in Fourier space is simply $\rho_k = \omega_k n_k/2$, where $\omega_k$ is defined via \eqref{omegaDef}. At early times, spectra must be plotted while $\phi_{(0)}<0$, because only during these intervals can one rotate fields into a basis that diagonalises the mass matrix $M_{ab}$. For positive scalaron, the expansion is more complicated due to the $SU(2)$ symmetry being broken by the scalaron background.

The simulation is stopped once the inhomogeneous energy fraction begins to grow slower than the Hubble rate, indicating the end of rescattering. Around this time, we find the energy in the Higgs to be dominated by its inhomogeneous modes, which  are relativistic and give a radiation-like equation of state. In the absence of further interactions, we may therefore analytically predict the evolution of the scale factor at later times by assuming the energy in homogeneous scalaron and inhomogeneous modes scale as $a^{-3}$ and $a^{-4}$ respectively. The radial Higgs mass is also obtained by scaling the homogeneous and inhomogeneous contributions to \eqref{mhDef} as $a^{-3/2}$ and $a^{-2}$ respectively. Once the combined scalaron decay width from $\Gamma_\phi$ \eqref{GammaPhi} exceeds the Hubble parameter, the scalaron quickly decays.  After this moment, which we denote with an ``r'' suffix, the potential \eqref{simplePotential} is simply quartic in $|h|$, and radiation-dominated expansion is maintained right up to thermal equilibrium.  

We may therefore calculate the number of e-foldings $N_e (k)$ before the end of inflation that the CMB pivot scale $k/a_0 \approx 0.002$/Mpc exits the horizon. We consider the expression
\be\label{neDef}
\mathcal{H}_* = \frac{k}{a_0} \frac{a_0}{a_r} \frac{a_r}{a_e} e^{N_e (k)}
\ee
where the ``$*$'' suffix denotes the moment of horizon exit, and $a_0$, $a_e$ and $a_r$ correspond to the scale factors at present, at the end of inflation, and at the onset of radiation domination, respectively. One finds (see appendix \ref{appendixNe} for details)
\be \label{neCalc}
N_e = 59.015 - \frac{1}{4} \log \frac{ \rho_r}{\rho_e}  - \log \frac{ a_r}{a_e}
\ee
where $\rho_e$ and $a_e$ are the grid-average energy density and scale factor at the end of inflation respectively. 

 Note that during the interval between $t_d$ and $t_r$, we ignore the backreaction of homogeneous scalaron decays when evolving the scale factor and $m_{h,\mathrm{eff}}$ forward in time, effectively assuming instantaneous decay. However, $N_e$ is logarithmically sensitive to the exact moment of the scalaron decay, so this approximation does not significantly change any observables. We calculate the scalar spectral index $n_s$ and and tensor-to-scalar ratio $r$ by evaluating the potential slow-roll parameters at the moment of horizon exit,
 \be
 \epsilon_* \equiv    \frac{1}{2M_P^2} \l\frac{V_{\mathrm{inf}}'(\phi_{(0)})}  {V_{\mathrm{inf}}(\phi_{(0)}) } \r^2_{\phi_{(0)}=\phi_*}, 
 \qquad 
 \eta_* \equiv  
     \left|  \frac{\,\,V_{\mathrm{inf}}''(\phi_{(0)})}{M_P^2 V_{\mathrm{inf}} (\phi_{(0)})  }
\right|_{\phi_{(0)}=\phi_*},
 \ee
and using the well-known formulas $n_s = 1-6\epsilon_* + 2\eta_*$, $r=15\epsilon_*$ \cite{Gorbunov:2011zzc}. This approach also yields the approximate relation
 \be
n_s \approx 1 - \frac{8 (4 N_e + 9)}{\l 4 N_e + 3 \r^2}, \qquad
r \approx \frac{192}{\l 4 N_e + 3 \r^2}.
\ee

\section{Results}
\label{sec:results}
\begin{figure*}[t]
	\centering
	\includegraphics[width=0.5\linewidth]{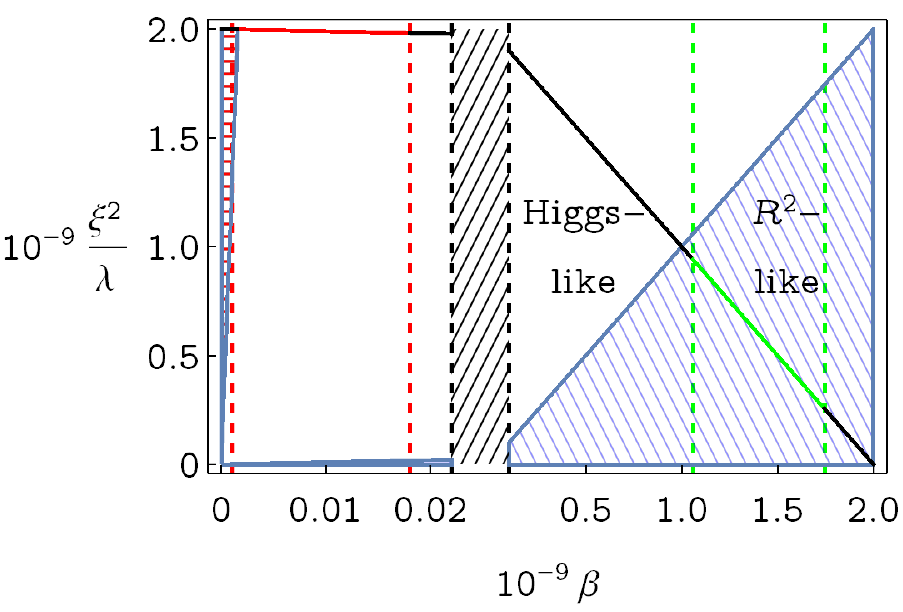}%
	\caption{
		Parameter space for the theory. The condition \eqref{conditon:CMB} constrains parameters to the solid line, and the red shaded region is excluded by \eqref{specNorm}. The white and blue regions correspond to ``Higgs-like'' and ``$R^2$ like'' parameters according to \eqref{conditon:CMB}. The red line corresponds to parameters studied in \cite{Bezrukov:2019ylq} and the green line corresponds to the parameters studied in this paper. Note the two horizontal linear scales for $0<\beta<2.5\times 10^7$ and $1.0\times10^8 < \beta< 2.0\times 10^9$: the solid straight lines in each region correspond to the same gradient and the black dashed lines indicate the region boundaries.
	}
	\label{fig:parameter_space}
\end{figure*}

As discussed in section \ref{sec:method}, our methods can be used for the $R^2$-like parameters $1.1\times 10^9 \lesssim \beta \lesssim 1.8 \times 10^9$, where the zero point energy is under control and reheating is complete before spectra becomes highly energetic and requires very small lattice spacing. The region of validity for a semiclassical treatment is shown inside the full parameter space in figure \ref{fig:parameter_space}.
In order to analyse the reheating for our parameter range, we run simulations for six values of $\beta$: $(\beta_1,\ldots ,\beta_6)\equiv(1.060, 1.214, 1.519, 1.556, 1.708, 1.745)\times10^9$. 

\begin{figure*}[t]
	\centering
	\includegraphics[width=0.5	\linewidth]{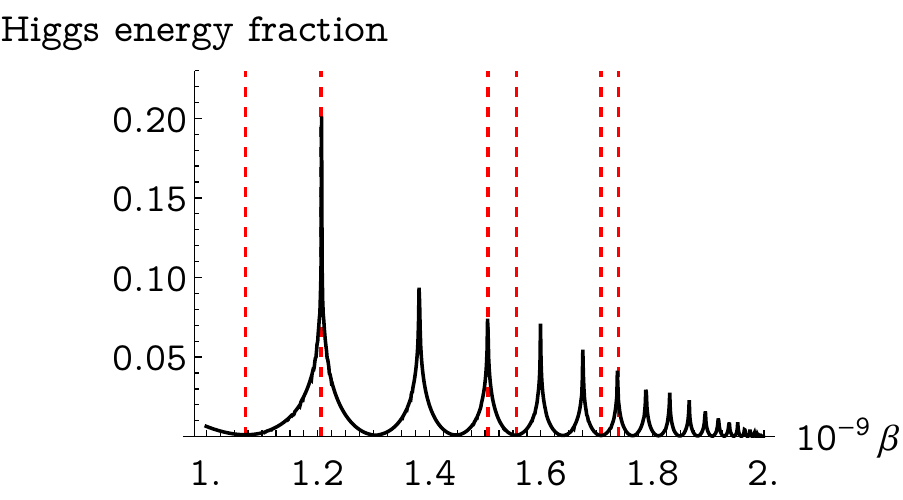}%
	\caption{
	Maximum kinetic energy fraction in the Higgs during the first scalaron oscillation for homogeneous dynamics only. The dotted red lines correspond to the parameters $\l \beta_1,\ldots,\beta_6\r$.
	}
	\label{peakPlot}
\end{figure*}

Initially critical dynamics transfers maximal energy from the scalaron to the Higgs after one scalaron crossing, if only homogeneous background evolution is analysed, and the variation of the transferred energy is plotted in figure \ref{peakPlot}.
We recognise $(\beta_1 \ldots \beta_6)$ as three pairs of adjacent maxima and minima for the Higgs energy fraction, corresponding to initially critical and noncritical dynamics respectively --- although with a maximum tuning of order $10^{-2}$ times the peak-to-peak distance once fluctuations are initialised. The qualitative variation of the preheating dynamics with $\beta$ can be understood by comparing the results of the initially-noncritical Higgs-like $\beta=\beta_1$ with the initially-critical $R^2$-like $\beta=\beta_6$.  We use the results of these two simulations, which we perform with $N=64$, $2 N \pi / L = 1000 m_\phi^{(+)}$ and an initial UV cutoff $\Lambda_{\mathrm{init}} =125m_\phi^{(+)} $, throughout this section. 

\begin{figure*}[t]
	\centering
	\includegraphics[width=0.5\linewidth]{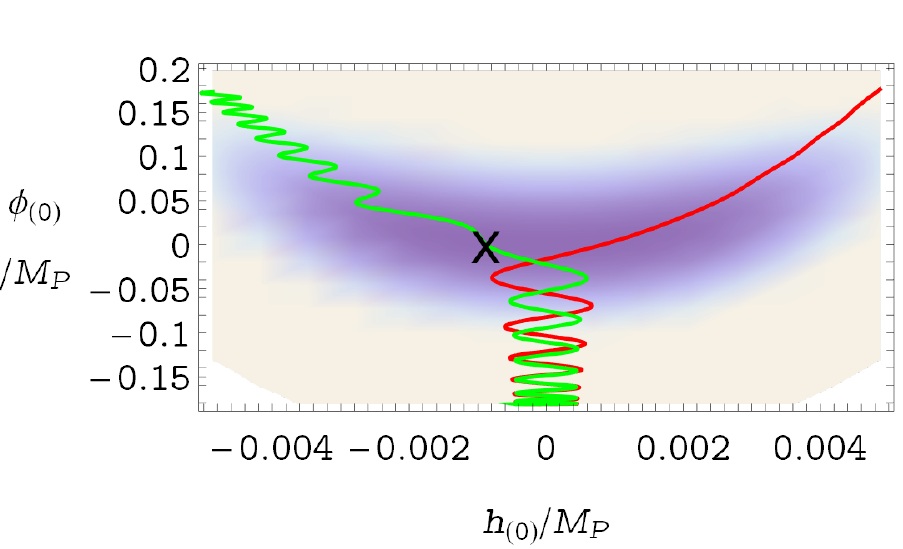}%
	\includegraphics[width=0.5\linewidth]{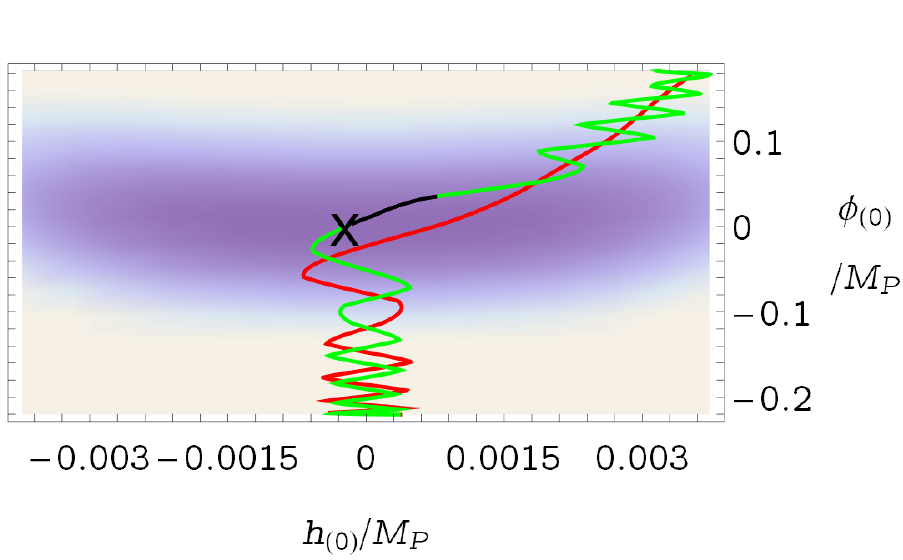}%

	\caption{
		Initial trajectories of the means $\phi_{(0)}$ and $h_{(0)}$, for $\beta=\beta_1$ (left) and $\beta=\beta_6$ (right). The red line corresponds to negative $\dot{\phi}_{(0)}$ and the green line corresponds to positive $\dot{\phi}_{(0)}$. Darker shading corresponds to a lower potential $V(\phi_{(0)}, h_{(0)})$. The black crosses correspond to the moment $\phi_{(0)}=0$, and the black line in the right hand plot corresponds to times with tachyonic $m_{h,\mathrm{eff}}^2$.
	}
	\label{trajectoryPlots}
\end{figure*}

\begin{figure*}[t]
	\centering
	\includegraphics[width=0.5\linewidth]{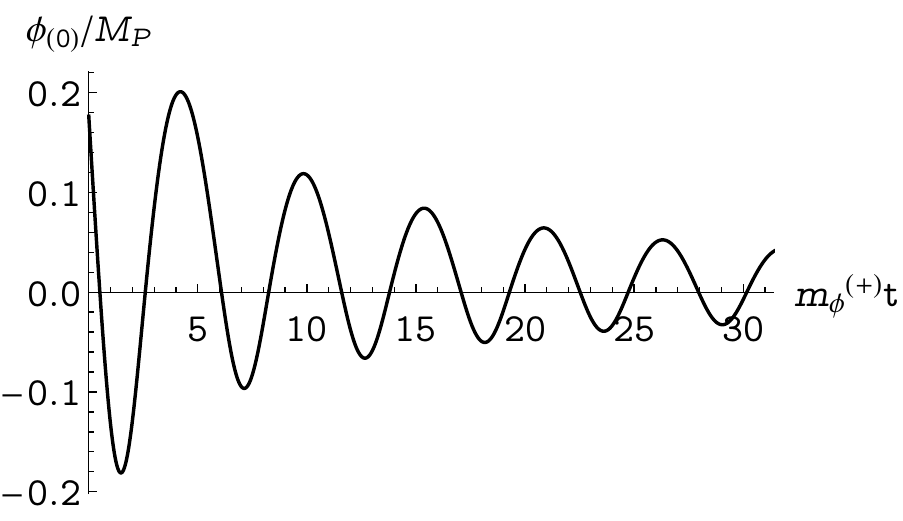}%
	\includegraphics[width=0.5\linewidth]{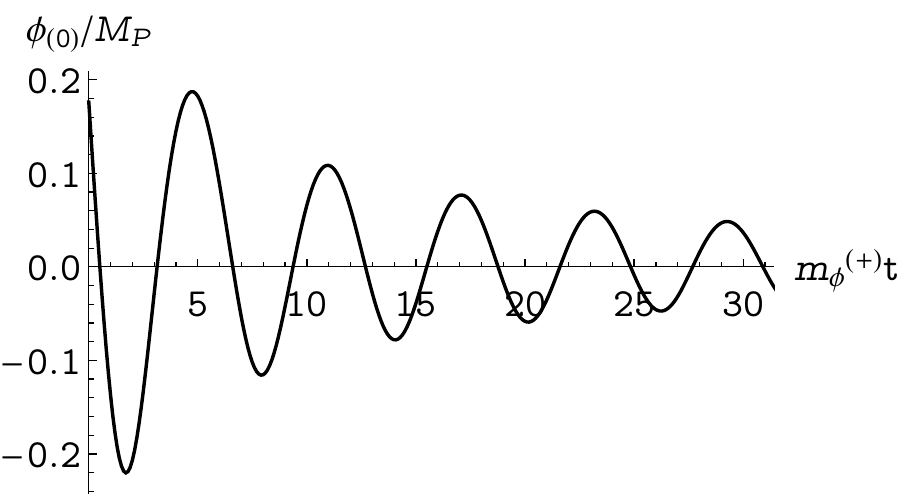}%
	\newline
	\includegraphics[width=0.5\linewidth]{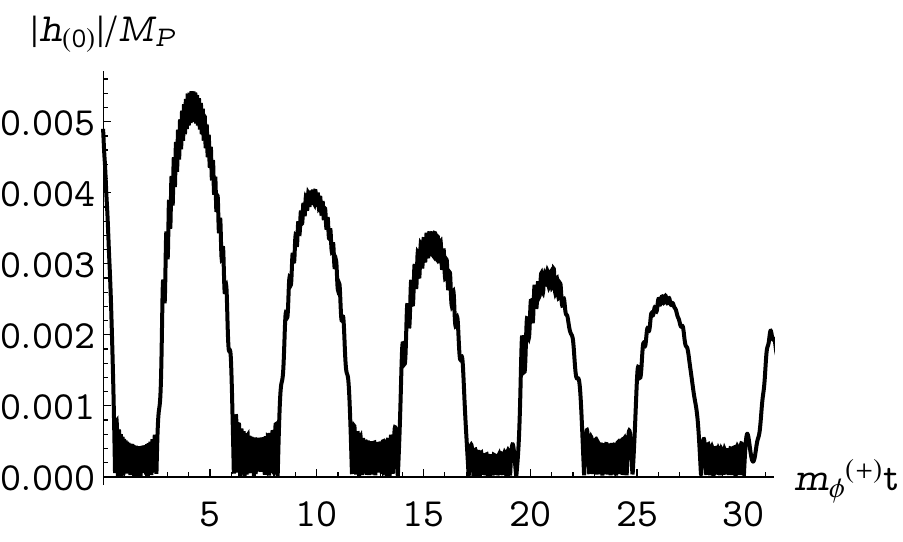}%
\includegraphics[width=0.5\linewidth]{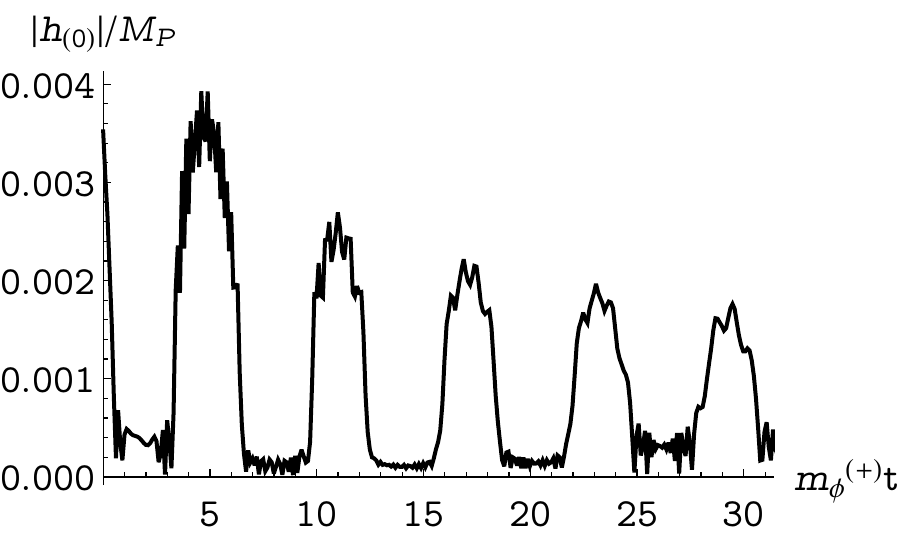}%
	\caption{
		Plots of the means $\phi_{(0)}$ (top), and $\left< h_1\right>$ (bottom, solid) and $\left< h_2\right>$ (bottom, dashed), for the first six scalaron oscillations, for $\beta=\beta_1$ (left) and $\beta=\beta_6$ (right).
	}
	\label{meansPlots}
\end{figure*}

\begin{figure*}[t]
	\centering
	\includegraphics[width=0.5\linewidth]{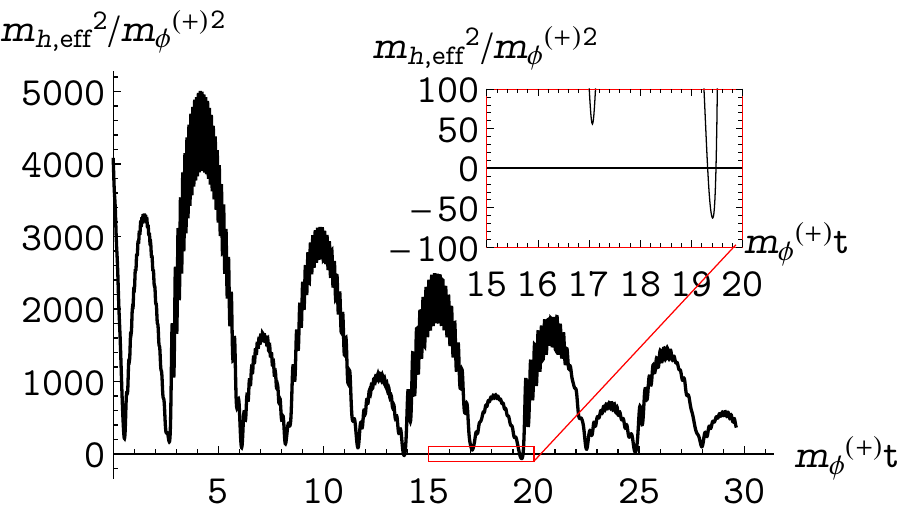}%
	\includegraphics[width=0.5\linewidth]{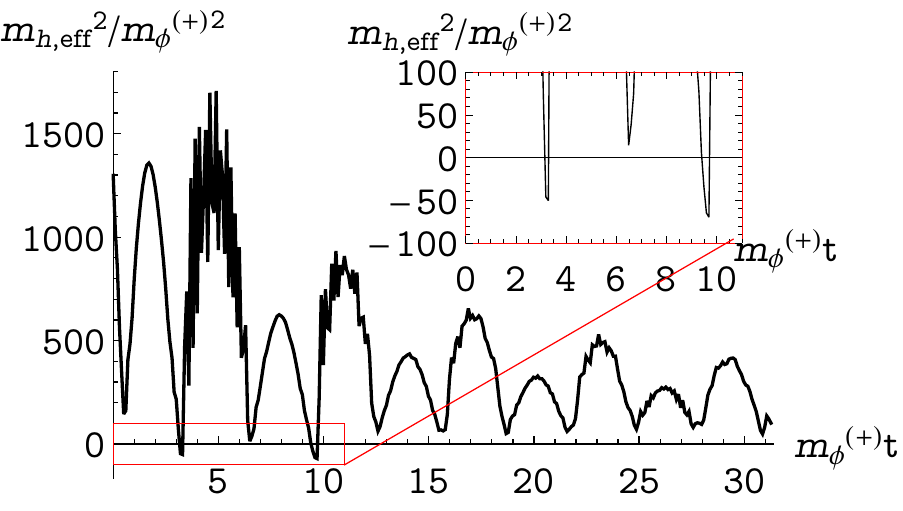}
	\caption{
	Plots of $m_{h,\mathrm{eff}}^2$ for the first six scalaron oscillations, for $\beta=\beta_1$ (left) and $\beta=\beta_6$ (right). $m_{h,\mathrm{eff}}^2$ becomes tachyonic after three scalaron oscillations for $\beta_1$, and immediately for $\beta_6$. The inset figures in the top right hand side of each plot zoom in on the boxed regions.
	}
	\label{mhPlots} 
\end{figure*}

Figure \ref{trajectoryPlots} shows the trajectories of the homogeneous fields $\phi_{(0)}$ and $h_{(0)}$ around the first scalaron crossing during the first scalaron oscillation.
Figure \ref{meansPlots} shows $\phi_{(0)}$ and $|h_{(0)}|$ plotted against time for the same parameter, during the first six scalaron oscillations.
When the scalaron crosses from negative to positive values during critical dynamics, the homogeneous Higgs spends longer around zero before dropping down to $h_{\mathrm{min}}(\phi_{(0)})$ than it does in the noncritical case.  While this additional time is a small fraction of a scalaron period, it is sufficient to induce a tachyonic instability in $m_{h,\mathrm{eff}}^2$, which is visible in figure \ref{mhPlots} where $m_{h,\mathrm{eff}}^2$ is calculated using \eqref{mhDef} and plotted against time.
All simulations performed give a stage of tachyonic resonance, which is realised within a few scalaron oscillations even for initially-noncritical dynamics, and lasts between one and two e-foldings. Although the homogeneous dynamics of $\beta=\beta_6$ were tuned to be maximally critical at the first scalaron crossing, they become somewhat detuned due to zero-point fluctuations and become more critical after one scalaron oscillation. After the initial tachyonic initial stage, the backreaction becomes sufficiently strong that the adiabaticity parameter
\be \label{adDef}
	\mathcal{A}\equiv\left| \frac{\dot{m}_h,\mathrm{eff}}{m_{h,\mathrm{eff}} ^2}\right|,
\ee
is less than one at all times. This terminates all non-adiabatic particle production and preheating continues via rescattering alone. 

\begin{figure*}[t]
	\centering
	\includegraphics[width=0.33\linewidth]{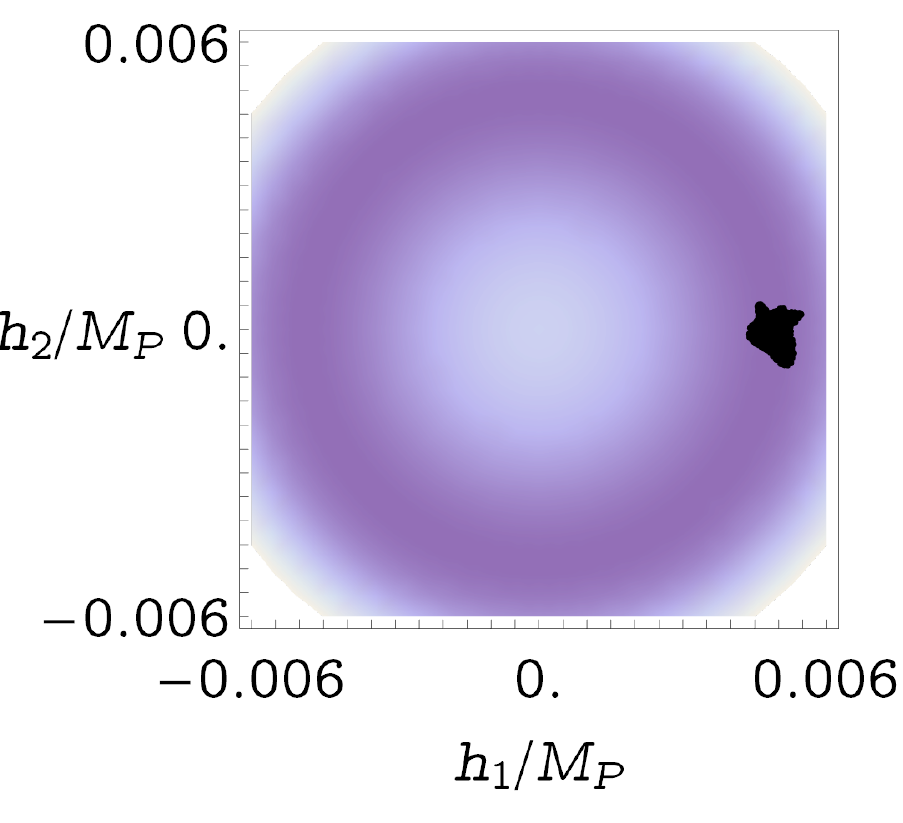}%
	\includegraphics[width=0.33\linewidth]{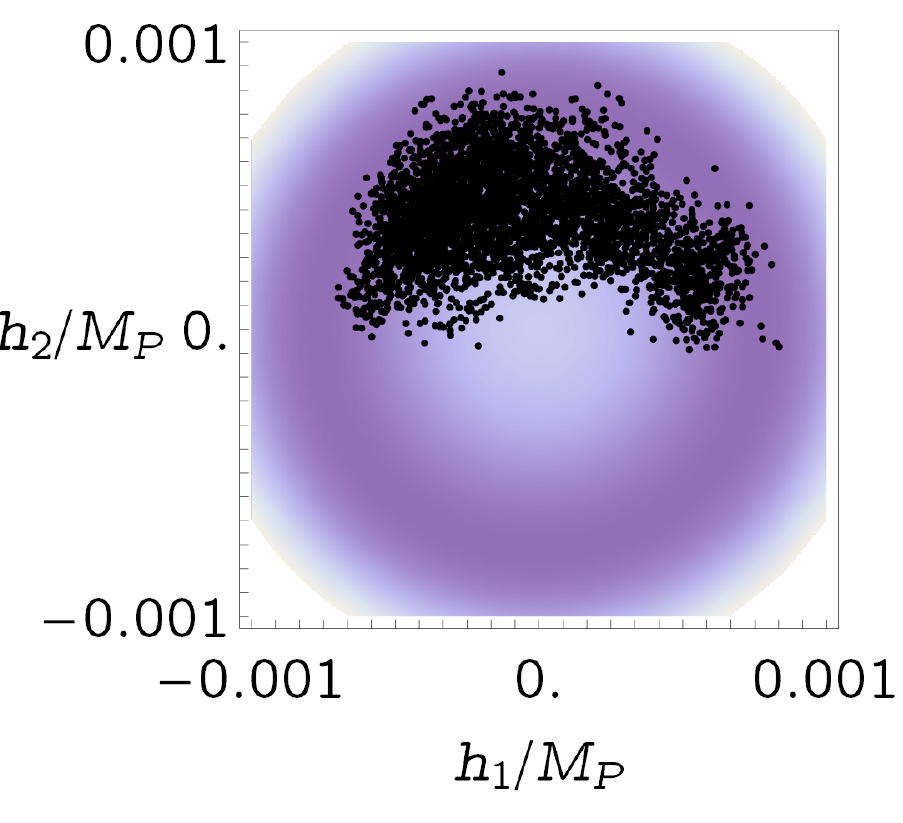}%
	\includegraphics[width=0.33\linewidth]{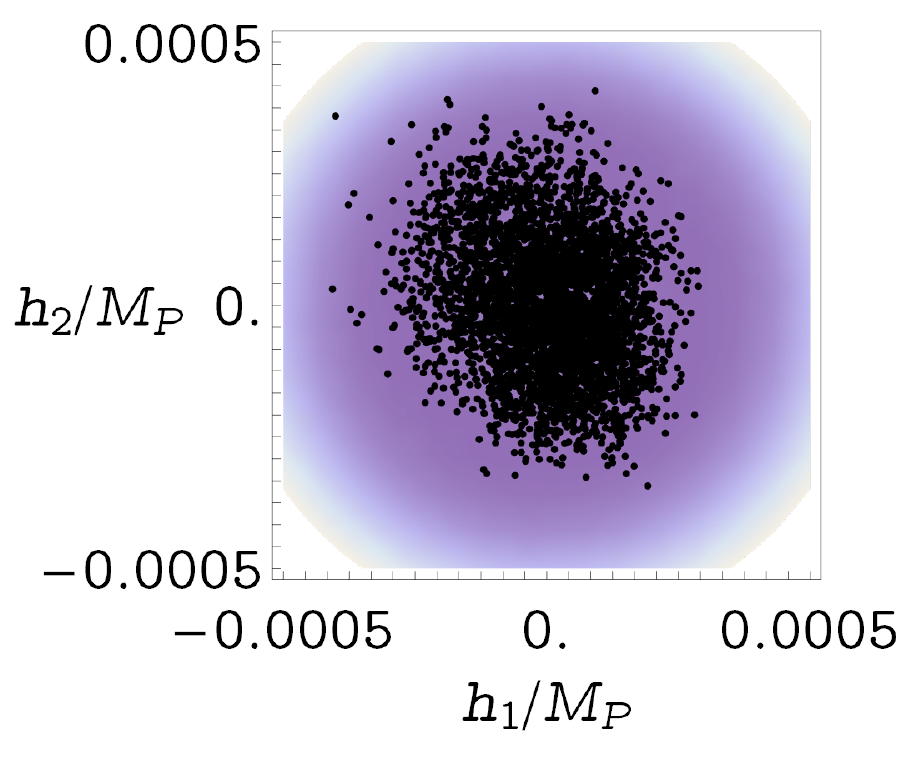}%
	\caption{
		Scatter plots of the  fields $h_1$ and $h_2$ at ${m_\phi^{(+)}} t = (0, 201, 501)$ (left to right) from all spatial points on a two-dimensional slice of the lattice. $\phi_{(0)}$ is maximally positive in its oscillation at each of these times. Darker shading corresponds to a lower potential $V(\phi_{(0)}, h_1, h_2, 0, 0)$.
	}
	\label{scatterPlots}
\end{figure*}

Figure \ref{scatterPlots} shows scatter plots of the fields $h_1$ and $h_2$ from all spatial points on a two-dimensional slice of the lattice for $\beta=\beta_1$ during moments when $\phi_{(0)}$ is maximally positive in its oscillation. Also plotted is the  potential $V(\phi_{(0)}, h_1, h_2, 0, 0)$.
At ${m_\phi^{(+)}} t=0$, rotational symmetry is broken by an expectation value in the $h_1$ direction and the other three Higgs directions are the Goldstone modes. At ${m_\phi^{(+)}}t=201$, the expectation value has rotated somewhat but the symmetry is still broken. By ${m_\phi^{(+)}}t=501$, symmetry is restored and the four Higgs directions become indistinguishable.

\begin{figure*}[t]
	\centering
	\includegraphics[width=0.5\linewidth]{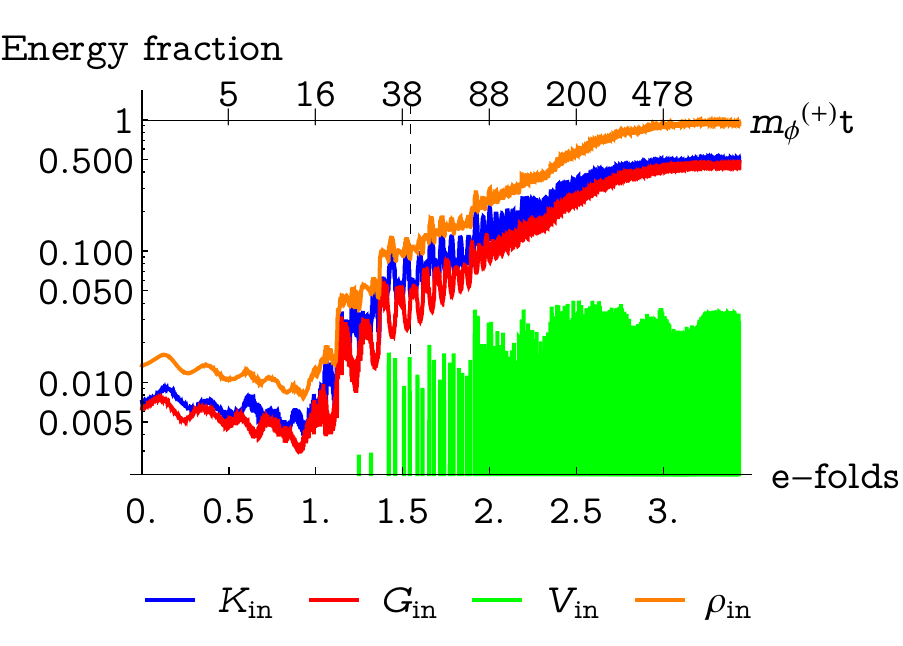}%
	\includegraphics[width=0.5\linewidth]{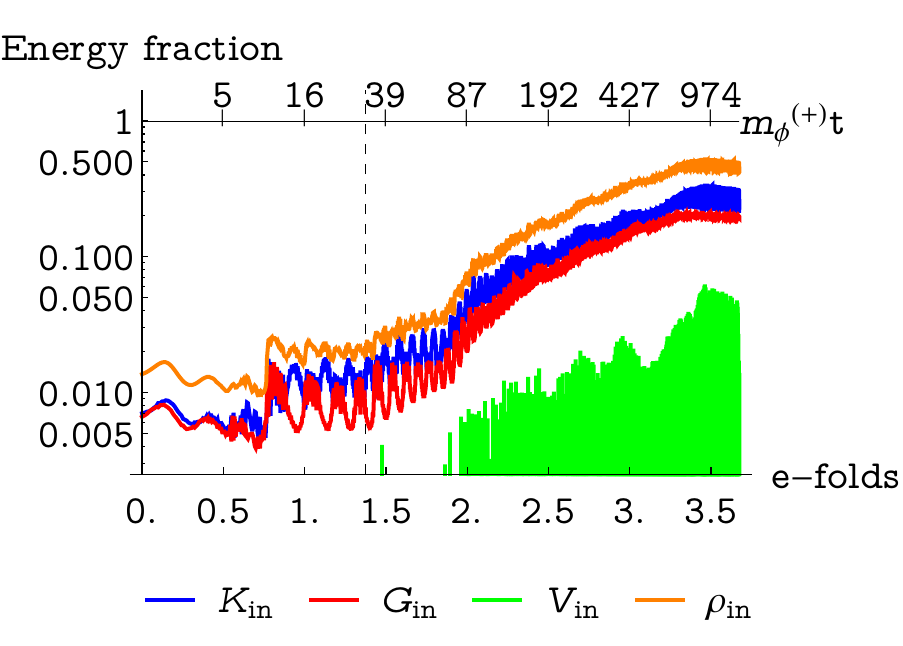}%
	\caption{
		Plots of the inhomogeneous energies $K_{\mathrm{in}}$, $G_{\mathrm{in}}$ and $\rho_{\mathrm{in}}$ with number of e-foldings for $\beta=\beta_1$ (left) and $\beta=\beta_6$ (right). The total energy is normalised to unity and the dashed vertical lines correspond to the last moment that the adiabaticity parameter $\mathcal{A}$ exceeds one.
	}
	\label{energyPlots}
\end{figure*}

Figure \ref{energyPlots} shows $K_{\mathrm{in}}$, $G_{\mathrm{in}}=G$ and $\rho_{\mathrm{in}}$, as defined in \eqref{kinDef}, plotted against number of elapsed e-foldings after the end of inflation, where the total energy is normalised to unity. 
The vertical dashed line denotes the moment that the adiabaticity parameter $\mathcal{A}$ drops to below one at all times. After this moment, the scalaron is further depleted by rescattering alone.

Comparing figures \ref{scatterPlots} and \ref{energyPlots}, preheating is nearly complete before the full restoration of symmetry. Between $\beta=\beta_1$ and $\beta=\beta_6$, rescattering becomes insufficient to completely destroy the scalaron. This is marked by a decrease in the inhomogeneous energy fraction at late times (around $3.5$ e-foldings for $\beta=\beta_6$), indicating the end of efficient rescattering. Specifically, once the Hubble rate exceeds the rescattering rate, modes with small $k$ are redshifted below the IR resolution $2\pi/L$ faster than new fluctuations are produced.

\begin{figure*}[t]
	\centering
	\includegraphics[width=0.5\linewidth]{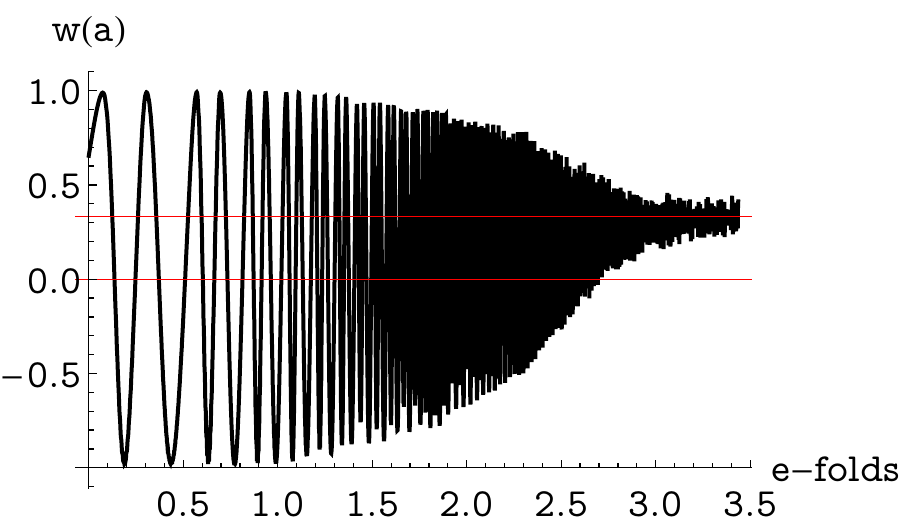}%
	\includegraphics[width=0.5\linewidth]{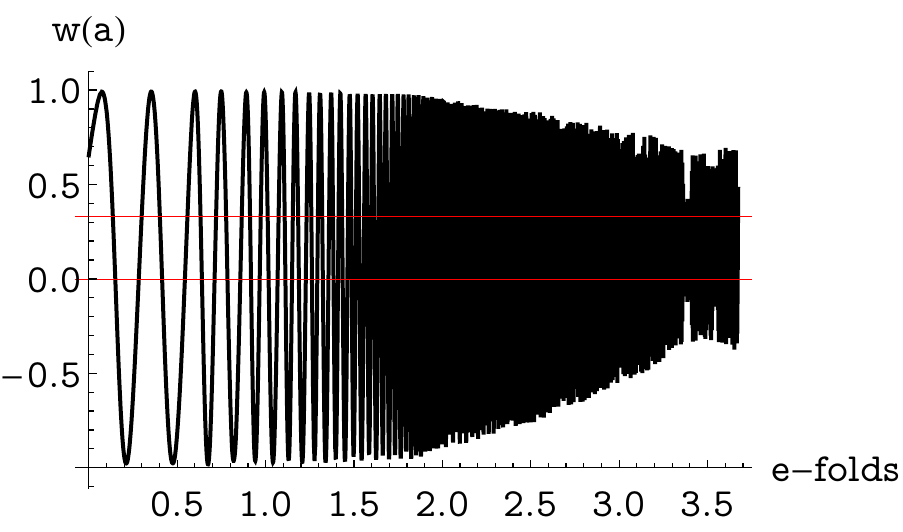}%
	\caption{
		Plots of $w(a)$ against number of e-foldings for $\beta=\beta_1$ (left) and $\beta=\beta_6$ (right). The two horizontal lines correspond to $w=0$ and $w=1/3$, for matter and radiation-dominated expansion respectively.
	}
	\label{expansionPlots}
\end{figure*}

For both $\beta=\beta_1$ and $\beta=\beta_2$, the profile of $G_{\mathrm{in}}/\l3 M_P ^2\mathcal{H}^2\r$ at late times shows that the energy of the system is dominated by relativistic degrees of freedom. This is confirmed in figure \ref{expansionPlots}, which plots the equation of state
\be
 w(a) \equiv -\frac{1}{3}	\frac{d \log \mathcal{H}^2}{d \log a}-1
\ee
against number of e-foldings since the end of inflation, over the same time range as \ref{energyPlots}.
Expansion begins with $w(a)=0$ for matter domination, and moves towards $w=1/3$ for radiation domination. 

\begin{figure*}[t]
	\centering
	\includegraphics[width=0.5\linewidth]{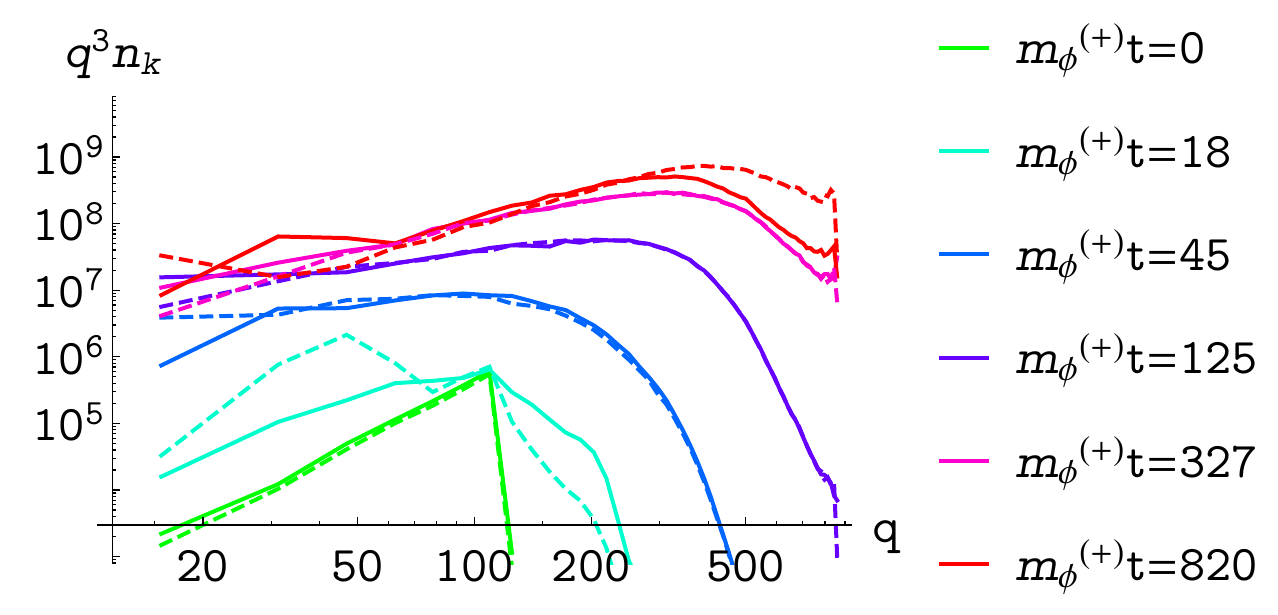}%
	\includegraphics[width=0.5\linewidth]{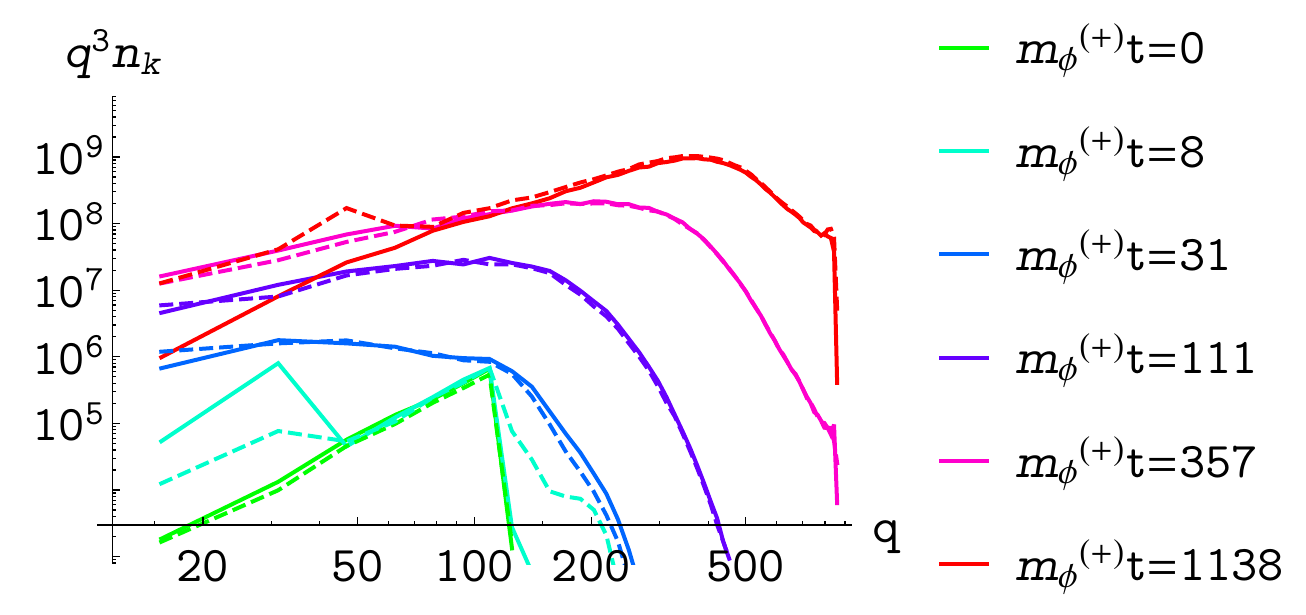}%
	\caption{
		Spectra of $q^3 n_\mbf{k}$ for the diagonalised fields approximately corresponding to $h_1$ (solid) and $h_2$ dotted, where the physical momentum $q$ is in units of $m_\phi^{(+)}$. The left hand plot has $\beta=\beta_1$ and the right hand plot has $\beta=\beta_6$. The times shown have $\phi_{(0)}<0$ and are separated by a constant logarithmic interval.
	}
	\label{spectraPlots}
\end{figure*}

Figure \ref{spectraPlots} shows the contribution of excited modes to energy $n_k\l\, k/a(t)\,\r^3$  plotted against the physical momentum $q(t)\equiv\l\, k/a(t)\,\r$, in the two of the orthogonal directions $\tilde{f}_1$ and $\tilde{f}_2$ that diagonalise the mass matrix $M_{ab}$ as defined in \eqref{mijDef}. These two directions correspond to  $h_1$ and $h_2$ respectively at the end of inflation. 
Early times are characterised by a cutoff in $n_\mbf{k}$ at momenta comparable to the maximally tachyonic mass, although somewhat greater at energy is efficiently transferred towards the UV almost immediately. Once non-adiabatic production stops, total occupation numbers continue to increase due to rescattering and the distribution cutoff moves towards the UV due the $|h|^4$ interaction in the regime of free turbulence. Once the global $SU(2)$ symmetry is restored, the the four Higgs directions become indistinguishable from one another. This is visible in figure \ref{spectraPlots}, where the spectra for the two Higgs directions coalesce at late times.

\begin{figure*}[t]
	\centering
	\includegraphics[width=.6\linewidth]{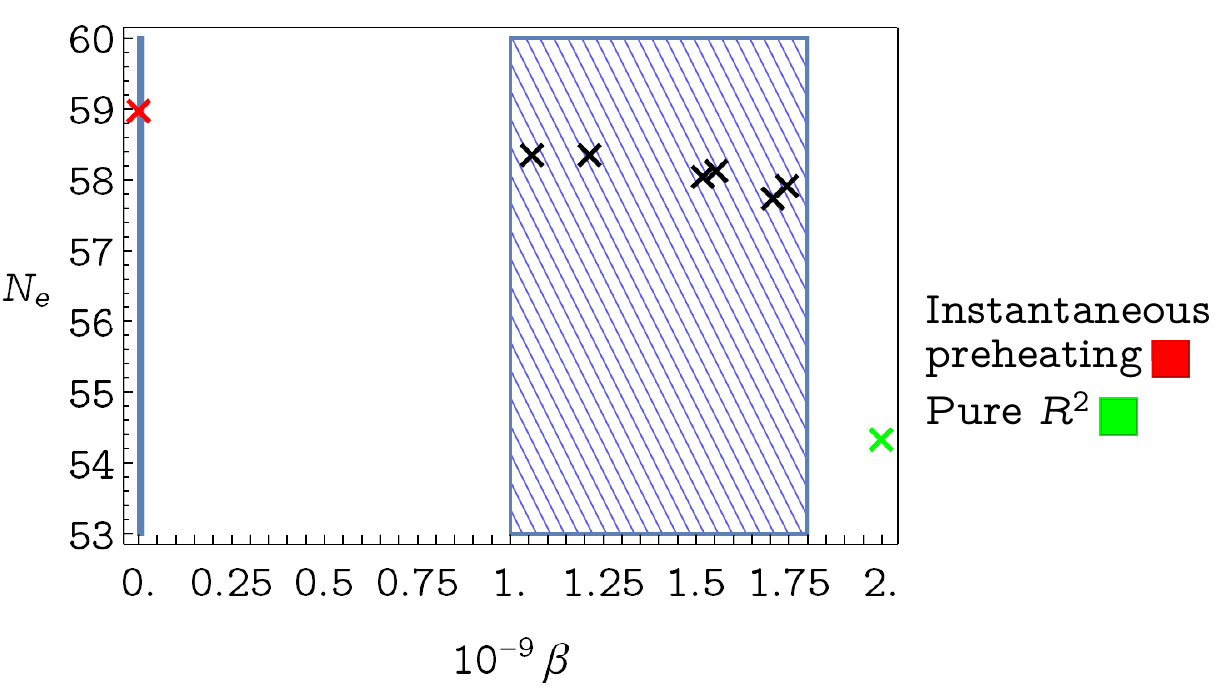}%
	\caption{
	Plot of the number of e-foldings $N_e$ against $\beta$ for the parameters that permit a lattice simulation (blue region) and for instantaneous (red) and $R^2$ (green) preheating. The narrow region on the left hand side is excluded from the parameter space by \eqref{conditon:perturbativity}. 
	}
	\label{neFig}
\end{figure*}

The scale factor is evolved beyond the simulation run-time as discussed in the previous section, and \eqref{neCalc} is used to calculate the number of e-foldings $N_e$ defined in \eqref{neDef} which we plot against $\beta$ in figure \ref{neFig}. 
Note that there is no clear pattern for the variation of $N_e$ depending on whether dynamics are initially critical or not. This is because results are mildly sensitive the box size and initial UV cutoff, which are chosen to compromise between a complete spectral profile and a manageable zero-point energy in fluctuations. There is a degree of arbitrariness in this compromise, and within this tolerance one cannot discriminate values of $\beta$ that give initially critical dynamics from those that do not.

\section{Discussion}
\label{sec:discussion}
Our predictions for $N_e$ in the mixed case lie between $N_e \approx 59$ and $N_e = 54.37$ for instantaneous and $R^2$ preheating respectively, where the latter is a result of \cite{Bezrukov:2011gp}, despite the particle production mechanism being qualitatively multifield in nature. Both the tachyonic mass and the rescattering controlled by $\xi$ appear exclusively in the mixed scenario, while in the $R^2$ limit the scalaron must decay through the kinetic terms of \eqref{scalarAction}, with the decay width \cite{Gorbunov_2011}
\be \label{gammaPhiDef}
\Gamma_\phi|_{R^2} = \frac{M_P}{144 \sqrt{6} \pi^2 \beta^{3/2}}
\ee
which is much smaller than $\Gamma_\phi$ for the mixed case.  Meanwhile, the preheating dynamics for pure Higgs inflation is strongly coupled, as discussed in section \ref{sec:intro}. The predictions for $N_e$ in the mixed case lie between the predicted values for $R^2$ and instantaneous preheating, where one expects the latter to approximate the pure Higgs scenario. One may investigate the pure $R^2$ by taking the $\xi \rightarrow 0$ limit of the action \eqref{scalarAction} and performing a perturbative study including scalaron decays. However, one cannot take the $\beta \rightarrow 0$ limit of our theory without violating the perturbativity condition \eqref{conditon:perturbativity}. 

\begin{figure*}[t]
	\centering
	\includegraphics[width=.4\linewidth]{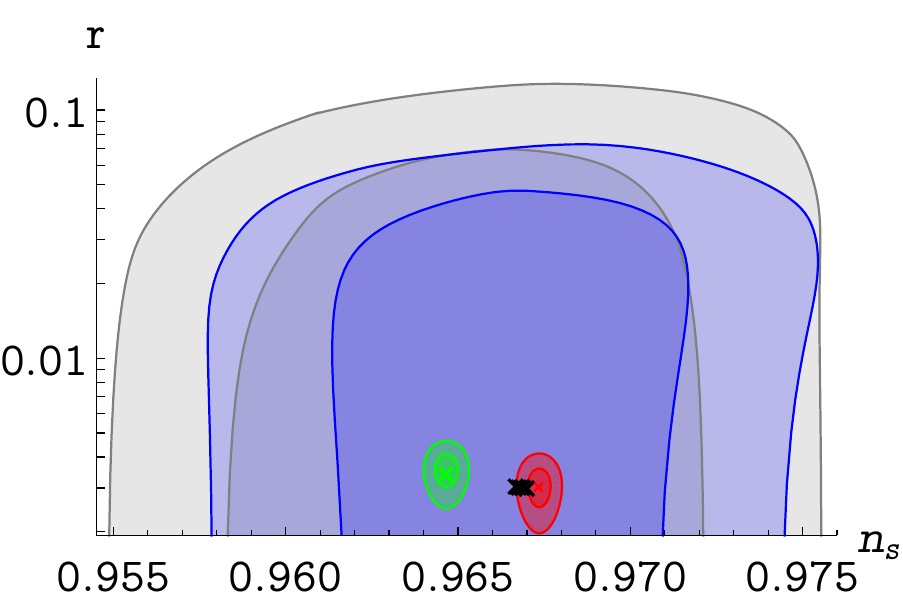}%
	 	\includegraphics[width=.6\linewidth]{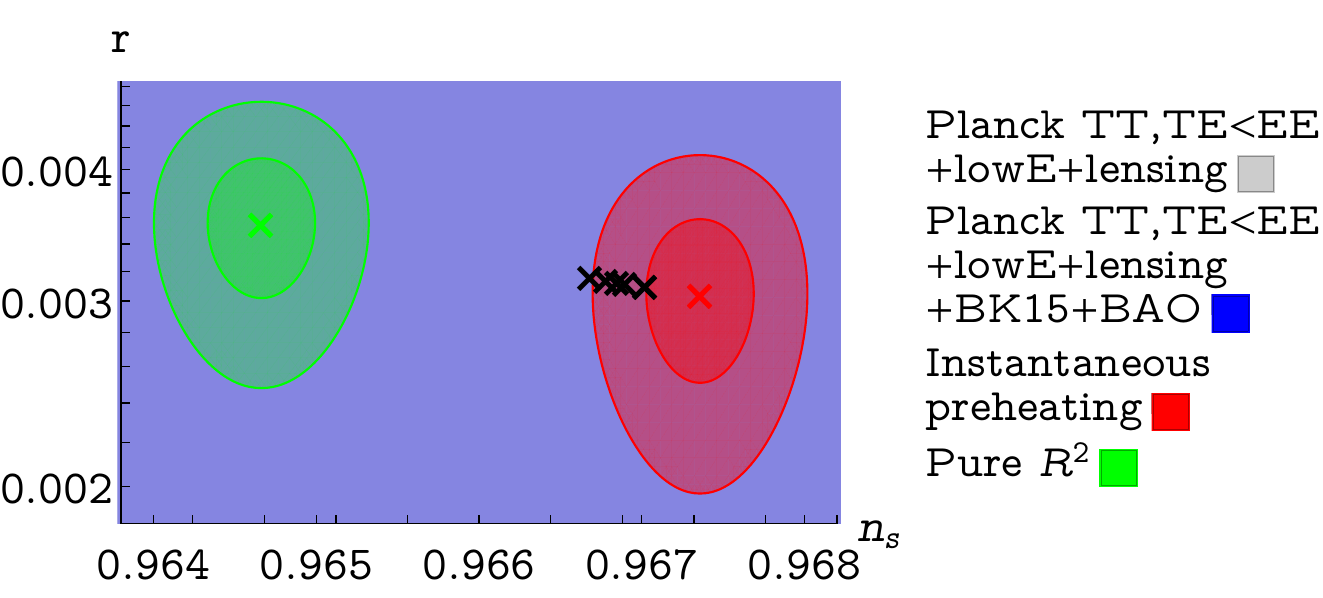}%
	\caption{
	Plots of the spectral index $n_s$ and the tensor-to-scalar ratio $r$ for the instantaneous and $R^2$ preheating, denoted by red and green markers respectively. The ellipses around the points correspond to the predicted $1\sigma$ and $2\sigma$ confidence intervals from our measurements. We use the uncertainty  $\Delta n_s = 0.0033$ for the FFTT experiment and $\Delta r = 5.4\times 10^{-4}$ for the EPIC-2m experiment. The black markers correspond to all six predictions for $n_s$ and $r$ from our simulations.  Also plotted are the $1\sigma$ and $2\sigma$ confidence intervals for the results of the Planck experiment. The right hand plot is a zoomed-in version of the left hand plot.
	}
	\label{fig:indexPlots}
\end{figure*}

In the region where semiclassical lattice methods can be readily used, all predictions for the spectral index $n_s$ and tensor-to-scalar ratio $r$ lie within the $68\%$ confidence interval of the Planck measurements $n_s=0.9649\pm 0.0042$, $r<0.056$ \cite{Akrami:2018odb}, and therefore may not be discriminated from one another at the time of writing. This situation is not improved by the forecast for the LiteBIRD experiment proposal \cite{Matsumura:2013aja}, which aims to measure $r$ with the $68\%$ confidence interval $\Delta r< 0.01$. A stronger discrimination using $n_s$ may be provided by future 21 cm tomography experiments such as the Fast Fourier Transform Telescope (FFTT), which offers the $1\sigma$ interval $\Delta n_s = 0.0033$ for realistic modelling of astrophysical contaminants \cite{Mao_2008}. This accuracy is far superior to hypothetical sky surveys such as the EPIC-2m experiment described in \cite{Baumann_2009} which offers $\Delta n_s = 0.0016$, $\Delta r = 5.4 \times 10^{-4}$, and would allow one to confidently discriminate between instantaneous preheating and the pure $R^2$ theory. This situation is displayed in figure \ref{fig:indexPlots}, where $n_s$ is plotted against $r$ for the instantaneous and $R^2$ preheating, as well as the results from our simulations, using $\Delta n_s$ for the FFTT and $\Delta_r$ for EPIC-2m. These predictions are shown inside the $1\sigma$ and $2\sigma$ regions of the Planck 2018 results.
While our confidence intervals may allow the $R^2$ theory to be discriminated from instantaneous preheating in the future, our predictions for the mixed case all lie within a $1\sigma$ interval both in $n_s$ and $r$, and may therefore not be discriminated from one another. 

We may extend our results to the Higgs-like parameters $\beta < 1.1\times10^9$ by observing that this region of parameter space is characterised by a larger tachyonic mass scale and a stronger coupling in the third term of \eqref{simplePotential}. These features give stronger tachyonic production and faster rescattering respectively. Although we can not directly study this region with semiclassical methods, we expect the preheating to be even faster than it is for $\beta=\beta_1$ --- which is to say, cosmologically near-instantaneous. 

The upper value of $\beta$ for our range is already close to the pure $R^2$ limit. The interpolation of our results to the pure $R^2$ case is nontrivial, as between $\beta \approx 1.8 \times 10^9$ and  $\beta=2.0 \times 10^9$ there is a region in parameter space where $m_{h,\mathrm{eff}} < {m_\phi^{(\pm)}}$ at all times after inflation. We expect the preheating in this narrow range of parameters to be qualitatively different to the more typical dynamics studied here, and we leave its investigation for a future publication.
 
Let us discuss the modification to our result from the inclusion of the gauge bosons.  Currently, we track only the actively produced Goldstone bosons, or, equivalently, the longitudinally polarised gauge bosons. The full physical preheating mechanism for our parameters also includes weak boson decays, whose decay width is
\begin{equation}
	\Gamma_W \simeq 0.8\, \alpha_W m_W,
\end{equation}
where  $\alpha_W \approx 0.025$ and $m_W$ is the $W^\pm$ boson square mass. During intervals with broken symmetry,
\be
	m_W^2 \approx \frac{g^2}{4} \left< |h|^2\right>,
\ee
where  $g\approx 0.56$ for the energy scales around the end of inflation. The situation is similar for $Z$ bosons. As far as $\Gamma_W$ is initially comparable with with the scalaron oscillation period, one may attempt to model weak decays by adding a frictional force in the Goldstone directions to \eqref{eom1} and \eqref{eom2} while $\phi_{(0)} >0$. However, such a force cannot discriminate between the broken and unbroken symmetry intervals, and would introduce unphysical dynamics to the latter. We therefore choose to omit $\Gamma_W$ entirely from our simulations.

This omission is justified by the following expectation regarding weak decays. The tachyonic stage is terminated by the growth of inhomogeneous fluctuations, and the depletion of the number of produced gauge bosons due to $\Gamma_W$ prolongs this stage. However, tachyonic production only lasts for a few scalaron oscillations, so we expect any extension to be a small fraction of the full physical time required for preheating.

We also omit perturbative Higgs decays, in order to fully preserve the comoving conservation of energy. The Higgs decay width $\Gamma_h$ in the unitarity gauge is dominated by decays into two $b$ quarks, and its value at rest in the unitarity gauge at zero temperature is
\begin{equation}\label{gammahDef}
\Gamma_h = 0.1 y_b^2 m_{h,\mathrm{eff}},
\end{equation}
where $y_b \approx 0.02$. This expression is valid until the full restoration of symmetry, which only occurs near the end of preheating. At early times, $\Gamma_h$ is around $10^{-1}$ times the Hubble parameter, and the radial Higgs cannot efficiently decay. At late times, $\Gamma_h$ is comparable to the Hubble parameter, but the Higgs oscillations are relativistic and the decays experience time dilation in the comoving frame by the usual Lorentz factor $\gamma$. Depending on the exact physical momentum chosen from figure \ref{spectraPlots}, the time-dilated decay width for fluctuations $\gamma^{-1} \Gamma_h$ is between  $10^{-2} $ and $10^{-3}$ times smaller than the Hubble parameter at all times, for all simulations performed. While one expects some deviation from \eqref{gammahDef} at late times with restored symmetry, this large disparity suggests that Higgs decays will be strongly subleading in the preheating dynamics right up to the onset of radiation-dominated expansion. 

\section{Conclusions} 
\label{sec:conclusions}
We have found that the complicated preheating of mixed Higgs-scalaron inflation reveals new intricacies when one attempts to move beyond a linearised treatment. Semiclassical lattice simulations can be performed for the parameters $1.1 \times 10^9 \lesssim \beta \lesssim 1.8 \times 10^9$. For these parameters, the critical dynamics that gives rise to tachyonic preheating occurs within a few scalaron oscillations, and whether or not dynamics are initially critical does not significantly affect observables. For smaller $\beta$ in our range the preheating is extremely fast, and thus for Higgs-like parameters $\beta< 10^9$, we expect observables to saturate the bound of instantaneous preheating.

The mixed preheating is characterised by a fast initial stage of tachyonic production, followed by a longer stage of rescattering and free turbulence. As one increases $\beta$ within the analysed range, one enters a situation where rescattering ends before the homogeneous scalaron is fully destroyed, and the scalaron decays perturbatively later. For our parameters, we find that the CMB pivot scale exits the horizon between $58.4$ and $57.8$ e-foldings before the end of inflation. This value may be slightly modified by including weak boson decays, which we expect to extend the tachyonic stage. These upper and lower values for $N_e$ give predictions of inflationary parameters $n_s=0.970 $, $r=0.0031$ and $n_s=0.9667$, $r=0.0032$ respectively. Measurements of the scalar spectral tilt from the proposed FFTT experiment, with the $1\sigma$ confidence interval $\Delta n_s = 0.0033$, may allow one to discriminate the mixed Higgs-$R^2$ theory from the instantaneous preheating and $R^2$ limits, but cannot strongly constrain the mixed parameter space.

\acknowledgments

The authors would like to thank Dmitry Gorbunov for thoughtful discussions during many stages of this work and Anna Tokareva for valuable comments. The work is supported in part by STFC research grant ST/P000800/1.

\appendix
\section{Calculating $N_e (k)$} 
\label{appendixNe}
Let us calculate the number of e-foldings $N_e (k)$ before the end of inflation that present-day momenta $k/a_0 \approx 0.002$/Mpc exit the horizon. We break up the period between horizon exit and the present day as
\begin{equation}
\mathcal{H}_* = \frac{k}{a_0} \frac{a_0}{a_r}\frac{a_r}{a_e} e^{N_e (k)}
\end{equation}
where the ``e'' and ``0'' subscripts denote the end of inflation and the present day respectively. Given that $\rho \propto a^{-4}$ after the scalaron is destroyed, one may define the reheating ``temperature''
\begin{equation}
\rho (t_r) = \frac{g_r \pi^2}{30} T_r^4
\end{equation}
where the ``r'' subscript denotes the onset of radiation-dominated expansion and $g$ is the number of effective relativistic degrees of freedom at a given time. Between the destruction of the scalaron and thermalisation, the conservation of energy gives $T \propto a^{-1}$, assuming $g$ does not drastically change. Once a thermal distribution is established, the comoving conservation of entropy
\be
	s = g \frac{2 \pi^2}{45} T^3
\ee
 gives $T \propto g^{-1/3} a^{-1}$, where $T$ is now a physical temperature. One may therefore write
\be
\frac{a_r}{a_0} = \l \frac{g_0}{g_r} \r^{1/3} \frac{T_0}{T_r}
\ee
which gives
\be
	N_e = \log \frac{T_0}{k/a_0} - \frac{1}{4} \log \frac{270}{\pi^2 g_0 ^{4/3}} - \frac{1}{12} \log g_r - \frac{1}{4} \log \frac{\rho_e}{\rho_*} - \frac{1}{4} \log \frac{\rho_r}{\rho_e} - \frac{1}{4} \log \frac{M_P ^4}{\rho_*} - \log \frac{a_r}{a_e}.
\ee
where a ``$\ast$'' suffix denotes the moment of horizon exit. The value of $\phi_e$ may be well approximated setting the slow-roll parameter \be
	\eta\equiv \left| \frac{V_{\mathrm{inf}}'' (\phi)}{M_P^2 V_{\mathrm{inf}} (\phi)} \right|_{\phi=\phi_e}
\ee
 equal to one, which gives $\rho_e \approx \l 1- \sqrt\frac{3}{4} \r^2 \rho_*$. With $g_0 = 43/11$, $g_r \approx 106.75$ (assuming no new physics up to the Planck scale) and $T_0 /q_0 \approx 10^{28}$,
\be
		N_e = 59.015 - \frac{1}{4} \log \frac{ \rho_r}{\rho_e}  - \log \frac{ a_r}{a_e}.
\ee
The sum of the last two objects is always negative; maximally so for long durations of reheating with matter-dominated expansion.

\bibliographystyle{utphys}
\bibliography{prelim_bib}

\end{document}